\UseRawInputEncoding

\documentclass[a4paper,11pt]{article}
\usepackage{pos}

\begin{document}

\title{Ultra-High Energy Proton-Proton Collision in the Laboratory System as the Source of Proton, Neutrino and Gamma Spectra in Astrophysics}

\author*[a]{Olga Piskounova}

\affiliation[a]{P.N. Lebedev Physics Institute,\\
  Leninski prosp. 53, Moscow, Russia}

\emailAdd{piskunovaoi@lebedev.ru}

\abstract{This paper is dedicated to the study how HE particle spectra, which are measured in cosmic ray physics and astrophysics, are influenced by the specifics of collider spectrum of protons. LHC experiments are providing us with the proton spectra at very high energy (VHE) measured in center-of-mass system, (c.m.s.). A QCD phenomenological study of previous years gave us the Quark-Gluon String Model for the modeling of baryon and meson production spectra in full kinematical range from centrally produced hadrons up to very forward ones. In 1990, I have applied the method for the recalculation the collider distribution of neutral pions into the description of gamma spectrum from supernova that was measured in the laboratory system. Here, I used the same method of spectra transfer to the laboratory system.  The main statement of this study is that the features of leading spectra of cosmic protons, neutrinos and gammas are to be formed already in first collision of UHE proton at the power source of cosmic particles. The specifics of collider spectrum of protons are as following: the growing distribution at the central rapidity’s, which reflects in the "knee" of cosmic proton spectrum, and the triple-Pomeron peak at the end of c.m.s. spectrum that leads to the enhancements in spectra of leading cosmic protons, neutrinos, gamma-photons, and as I assume, positrons too. These conclusions desire for more experimental facts to be collected as well as new observations to be made.
}

\FullConference{%
  41st International Conference on High Energy physics - ICHEP2022\\
  6-13 July, 2022\\
  Bologna, Italy
}


\maketitle

\section{Introduction}
The aim of this paper is to conclude on the proton spectrum, which initiates specific forms of UHE particle spectra in space \cite{partdata}. As it has been shown previously \cite{piskoun20},on the basis of LHC data at $\sqrt{s}$ = 7 TeV, that the knee in cosmic ray (CR) particle distribution is not the result of substantial change in the kinematics of hadron interaction. Here, we are studying, what form the proton spectrum will have in laboratory system after the first collision of HE protons that are injected from the UHE source of the cosmic particles.
I am intending to compare the form of CR particle spectrum with the result of Quark-Gluon String Model \cite{kaidalov} that was calculated for proton-proton collision at $\sqrt{s}$ = 540 GeV in c.m.s., which corresponds to the energy $1.6* 10^{5}$ GeV in the laboratory system.

\section{QGSM proton spectrum at the proton collision with the energy 540 GeV}
\begin{figure}[ht!]
\centering
\includegraphics[width=8.0cm, angle=0]{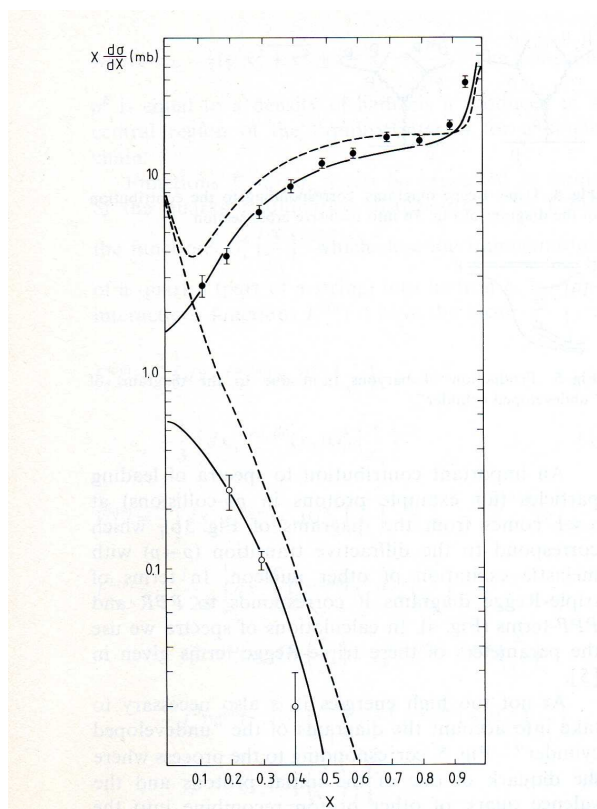}
 \caption{QGSM production spectra of protons in p-p collisions at colliders: the predictions for proton and antiproton spectra at the energy 540 GeV are shown with dashed lines, proton and antiproton distributions at lower energy are shown with solid lines.}
 \label{proton540}
\end{figure}
The spectra of protons from the collider data have been described in QGSM in 1980th years \cite{kaidalov}. It was already known about he growing density of particles at the central rapidity’s. This is vacuum pairs production and
the densities for protons and antiprotons are the same. The peak of proton production at the end of spectra was discussed already in \cite{pomeranchuk} in 1960th.

\section{Transfer to the laboratory system}

The procedure for recalculating of c.m.s. spectrum into the laboratory system one is easy. It was applied in the early description of gamma spectrum from supernova \cite{gamma}. The LHC experiments observed the central rapidity "table" in the spectra of produced hadrons. This spectrum should be a) shifted to the right by the value of $Y_{max}$ = ln(2$\sqrt{s}$/$m_p$), and b) divided per $E_{lab}$.

\section{Resulting proton spectrum in space}

\begin{figure}[ht!]
\centering
\includegraphics[width=12.0cm, angle=0]{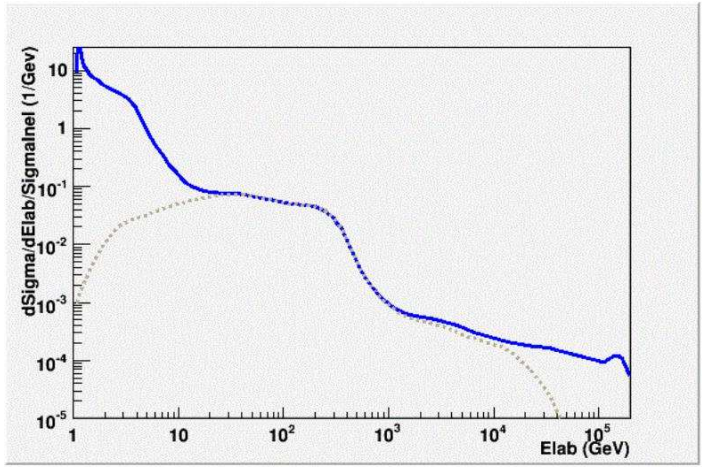}
 \caption{The form of proton spectra after UHE p-p collision in laboratory system, the antiproton spectrum is shown with dashed line.}
 \label{spaceprotons}
\end{figure}

The form of resulting differential proton spectrum is shown in the figure~\ref {spaceprotons}. In laboratory system, the central rapidity "table" gives us the "knee" and the triple-Pomeron peak corresponds to the bump at the end of CR proton spectrum. By the way, if we are taking the position of "knee" in the real cosmic proton spectrum, then the energy of initial proton should be of order $E_0 = 10^{10}$ GeV. 

\section{Spectra of neutrinos and gamma-photons} 

The bump at PeV energies is also seen in the spectrum of VHE neutrinos \cite{neutrinos}, which appear from the decays of heavy baryons as well. 
The gamma spectrum in the entire energy diapason, which was presented on the front cover of book \cite{gammabook}, also has a distinct bump that is a signature of accelerated proton collision.

\section{Conclusions}

Main conclusions are the following: 

i) Protons and antiprotons are the fundamental matter in the Universe. They cannot just annihilate. Their interactions are the source of many other particles in space: secondary protons and antiprotons, positrons and electrons, gamma-photons, and neutrinos.

ii) Proton spectrum in the laboratory system has the similar features as accelerator one: the central rapidity "table" is transfered to the knee and the triple-Pomeron peak corresponds to the bump at the end of CR proton spectrum.  It should be taken into account that every additional hadronic collision (or decay) brings the factor $E^{-1}$ into the resulting spectrum.

iii) The UHE protons are injected from the most powerful sources, Super Massive Black Holes (SMBH), and initiate CR particle spectra with distinctive signatures.

As a working hypothesis, I have made the assumption \cite{TorusDM} that DM is also built, in some symmetric way, from protons and antiprotons, which are compressed under extreme gravitation pressure at the BH’s.


\begin{thebibliography}{99}
\bibitem{partdata}M. Tanabashi et al. {\it Review of Particle Data Group}, Phys. Rev. D 98 (2018) 030001 
\bibitem{piskoun20}O.Piskounova, {\it Baryon production at LHC experiments: average pt of hyperons versus energy}, Int. Jou. of Mod. Phys. A {\bf 35} (2020) 2050067 [arXiv:1706.07648]
\bibitem{kaidalov}A.Kaidalov and O.Piskunova, {\it Inclusive Spectra of Baryons in the Quark-Gluon Strings Model}, Z. Phys. C {\bf 30} (1986) 145
\bibitem{pomeranchuk}V.N. Gribov, I.Ya. Pomeranchuk and K.A. Ter-Martirosyan, {\it Partial waves singularities near j=1 and high energy behaviour of the elastic scattering amplitudes}, Phys. Lett. {\bf 9} (1964) 269
\bibitem{neutrinos}M. S. Muzio, G. R. Farrar, and M. Unger, {\it Probing the environments surrounding ultrahigh energy cosmic ray accelerators and their implications for astrophysical neutrinos}, Phys. Rev, D {\bf 105} (2022)023022 [arXiv:2108.05512]
\bibitem{gamma}O.I. Piskunova, {\it Shape of gamma spectra from cosmic sources of high energy protons}, Sov. Journal of Nucl. Phys. {\bf 51} (1990) 1332
\bibitem{gammabook}F.A. Aharonian, {\it Very High Energy Cosmic Gamma Radiation, Crucial Window on the Extreme Universe}, World Scientific Publishing, 2004
\bibitem{TorusDM}O.I. Piskounova, {\it Baryon charge asymmetry at LHC: String Junction transfer in proton reactions and baryon-antibaryon SJ torus as DM candidate} [arXive:1812.02691]
\end{thebibliography}
\end{document}